\documentclass[12pt,preprint]{aastex}

\usepackage{amsmath}

\begin{document}

\title{Nonconservative Mass Transfer in Massive Binaries and the Formation of Wolf-Rayet+O Binaries
}
\author{
Yong Shao$^{1,2}$ and Xiang-Dong Li$^{1,2}$}

\affil{$^{1}$Department of Astronomy, Nanjing University, Nanjing 210046, China; shaoyong@nju.edu.cn}

\affil{$^{2}$Key laboratory of Modern Astronomy and Astrophysics (Nanjing University), Ministry of
Education, Nanjing 210046, China; lixd@nju.edu.cn}

\begin{abstract}

The mass transfer efficiency during the evolution of massive binaries is still uncertain.
We model the mass transfer processes in a grid of  binaries to investigate the
formation of Wolf-Rayet+O (WR+O) binaries, taking into account two kinds of
non-conservative mass transfer models: Model I with rotation-dependent mass accretion 
and Model II of half mass accretion.
Generally the mass transfer in Model I is more inefficient, with the average efficiency in
a range of $\sim0.2-0.7$ and $ \lesssim0.2 $ for Case A and Case B mass transfer, respectively.
We present the parameter distributions for the descendant WR+O binaries.
By comparing the modeled stellar mass distribution with the observed Galactic 
WR+O binaries, we find that highly non-conservative mass transfer is required.
%Moreover, there should be extra angular momentum loss mechanisms, e.g. outflow from 
%the outer Langrangian point, to account for their orbital evolution.

\end{abstract}


\section{Introduction}

Wolf-Rayet (WR) stars are massive helium burning stars, formed either through stellar wind mass
loss \citep[e.g.,][]{mm03,e08,sht12} or mass transfer in a binary system \citep[e.g.,][]{w01,p05}.
%Massive stars with mass $ \gtrsim 20-25 M_{\odot} $ will become WR stars due to a strong wind 
Observations show  that the majority of massive stars are in binary systems \citep{sd12,k14}.
The evolution of massive binaries plays a vital role in various aspects of astrophysics, and their
evolutionary products are related to many interesting phenomena, e.g. type Ib/c supernovae
\citep{p92}, high-mass X-ray binaries, and double compact star systems \citep{bv91}.
Mass transfer in a binary can dramatically influence the evolution of the system, changing
the properties of both stars and the binary orbit. During the process of Roche-lobe overflow (RLOF),
the primary star loses most of its hydrogen envelope, leaving a burning helium core, while mass accretion onto the secondary star causes it to rejuvenate and spin up
\citep{p81,h02}. After the mass transfer, the binary evolves to be a WR+O system.
However the mass transfer efficiency $ \beta $, i.e., the fraction of the transferred mass that is accreted
by the secondary, still remains uncertain.

Evolutionary calculations of massive binaries involving the formation of WR+O binaries
have been performed by many authors, since the pioneering works by \citet{p67}, \citet{hh72}, and \citet{v79}.
\citet{v82} considered the evolutionary scenario of massive close binaries with the primary mass
between $ 20M_{\odot} $ and $ 120M_{\odot} $, and showed that the mass
transfer efficiency $ \beta $ is required to be less than 0.3 in order to reproduce the
observations. \citet{dd92} investigated the evolutionary sequences of massive Case B binary systems
with primary masses $ 9-40M_{\odot} $ with the assumption of $ \beta = 0.5 $.
\citet{w01} presented the evolution calculations of massive binaries with conservative mass transfer,
they found that none of the observed WR+O binaries can fit the calculated results.
\citet{p05} explored the progenitor evolution of three WR+O binaries with the WR/O mass
ratio of $ \sim0.5 $ and the orbital periods of $ 6-10 $ days, and concluded that the mass transfer
must have be highly non-conservative. More recently,
\citet{d07} followed about $ 2\times10^{4} $ binary evolutionary tracks with
the primary  mass of $ 3.5-35M_{\odot}$ and the orbital period of $ 1-5 $ days. By comparing the non-conservative models tend to better match
the observed mass-transferring binaries in the Small Magellanic Cloud.  In our previous
work \citep{sl14}, we evolved a large number of binary systems and adopted three different
mass transfer models, from highly non-conservative to quasi-conservative, to derive the parameter
spaces of avoiding the occurrence of contact binaries. We showed that, an efficiency of 0.5 is required to be consistent with the observed  Be/X-ray binaries with a neutron star,  while the binary systems hosting a black-hole favors the less efficient mass transfer model.

In this work, we revisit the evolution of massive binaries. We attempt to use the  observed Galactic WR+O binaries to constrain the possible formation channels of these binaries, especially
the mass transfer efficiency. The remainder of this paper is organized as follows. In Section 2, We introduce the binary evolution code and the adopted assumptions. We present our calculated results in Section 3, and discuss their implications in Section 4.

\section{Binary Evolution Code and Method}

We use an updated version of the stellar evolution code
originally developed by \citet{e71,e72} \citep[see also][]{p95} to calculate the binary evolution.
We employ the TWIN mode in the code to model the structure and spin of both stars simultaneously
\citep{ye05}. We adopt the initial solar chemical compositions
(i.e., $X = 0.7$, $Y = 0.28$, and $Z = 0.02$), and the ratio of the mixing length to the pressure
scale height and the convective overshooting parameter to be 2.0 and 0.12 \citep{s97}, respectively. 
For the pre-WR stars, we employ the wind mass loss rates of \citet{dnv88}, except for OB stars
for which we adopt the modelled rates of \citet{vdl01}. For the WR stars [$ X_{\rm surface}<0.4 $,
$ \log(T/K)>4.0 $, $ \log(L/L_{\odot}) >5.0]$, we use the rates of \citet{nl00}. When all of hydrogen
envelope of the primary is stripped away by mass transfer and stellar winds, the code will be artificially
broken down.
The spin-orbit coupling due to tidal interactions is treated according to \citet{h81}.
The mass transfer rate via RLOF in the code is calculated from the potential difference
\begin{equation}
\frac{d\dot{M}}{dm}=-10^4\frac{\sqrt{2\Delta\phi}}{r},
\end{equation}
where $m$ and $r$ are the mass coordinate and the stellar radius,
respectively, by
integrating the above equation over all the mesh points outside the RL surface
potential.

The stability of the mass transfer  depends on how much and how fast the secondary can accrete
without getting out thermal equilibrium \citep[][and references therein]{h02}.
Unstable mass transfer would result in the formation of a
contact binary \citep{ne01}, which may finally evolve to a merger. Since accretion of a small
amount of mass can accelerate the star to reach critical rotation \citep{p81}, it is expected that,
during most of the mass transfer phase, the accreting star is rapidly rotating, and it is of vital
importance to determine the mass accretion efficiency for such a star
\citep{l98,p05,d09,se09}. Considering the large uncertainty in this point,
\citet{sl14} constructed three models to model the influence of mass loss on the stability
of mass transfer. In Model I, the mass accretion rate onto
a rotating star was assumed to be the mass transfer rate multiplying an ad hoc factor
$ (1-\Omega/\Omega_{\rm cr})$, where $ \Omega $ is the angular velocity of the secondary and
$ \Omega_{\rm cr} $ is its critical value. The stellar rotation was treated as rigid body, 
which controlled by spin up due to mass accretion and the spin-orbit coupling through tidal interactions;
in model II, the mass accretion efficiency was fixed to be $50\%$;
in Model III, the mass accretion rate was assumed to be limited by a factor of
$ \min (10\frac{\tau_{\dot{M}}}{\tau_{KH2}},1)$, where $\tau_{\dot{M}}$ denotes the mass transfer timescale 
and $\tau_{KH2}$ is the thermal timescale of the secondary \citep{h02}. 
As a result of rapid mass accretion, the secondary star would get out of thermal equilibrium, expand
and become overluminous. The thermal timescale of the secondary would be significantly decreased, so that
the mass transfer was generally quasi-conservative.
Since previous studies have already disfavored conservative mass transfer in massive binaries,
in present study we only consider Models I and II in our calculations.  We assume that the part
of the material that is not accreted by the secondary is ejected from the system in the form of
isotropic wind, taking away the specific angular momentum of the secondary.

\section{Binary Evolutionary Sequences}

We have calculated a series of binary evolutionary sequences with different values
of the initial parameters. The initial primary mass $ M_{\rm 1.i} $
(in units of $ M_{\odot}$), mass ratio $ q_{\rm i}= M_{\rm 1,i}/M_{\rm 2,i} $
and orbital period $ P_{\rm orb,i} $ (in units of days) are set as follow,
\begin{equation}
\begin{array}{cccccc}
 M_{\rm 1,i}   &=& 25, &30,& \ldots,& 60, \\
\log q_{\rm i}     &=& 0.1,& 0.2,&  \ldots, & 0.6, \\
\log P_{\rm orb,i} &=& 0.0, &0.1, &\ldots,& 2.0. \\
\end{array} \nonumber
\end{equation}
Both the binary components are zero-age main-sequence stars at the beginning of binary evolution.
If the initial orbital period is so short that the primary has filled its RL, we skip to the next
longer orbital period.

There are several ways to produce a WR+O binary. (1) If the initial orbital period is around a few days,
RLOF begins when the primary has a burning hydrogen core with Case A mass transfer. After
the main-sequence evolution, the primary experiences a rapid contraction due to the depletion
 of fuel in the convective core,  causing the binary to be detached. When the primary expands
due to shell hydrogen burning and fills its RL once more, Case AB mass transfer takes place.
After the mass transfer, the primary has lost most of its hydrogen envelope,
leaving a helium core burning WR star. At that time the secondary is a main-sequence O star
with an increased mass due to mass accretion.
(2) If the initial orbital period is several weeks, the primary is in the shell hydrogen burning phase
when RLOF starts. After this Case B mass transfer, the primary becomes a WR star.
(3) If the initial orbital period is of the order of years, there is Case C mass transfer
when the primary fills its RL during helium shell burning. The mass transfer proceeds on a
dynamical timescale, and common envelope evolution may follow, which will not be investigated
in this work.

Based on the spectra features, WR stars can be divided into four subtypes: the nitrogen sequence,
WNH (or WNh) and WN, with strong lines of nitrogen; the carbon sequence, WC, with strong lines of carbon;
the oxygen sequence, WO, with strong lines of oxygen. The H or h indicates WN stars with hydrogen. 
%The WN and WC stars present the products of the CNO cycle (H-burning)
%and the triple-$\alpha$ (He-burning), respectively. 
During the WR phase, the stellar mass  
decreases as a consequence of a powerful stellar wind \citep{h95,nl00}, and the masses of WC stars
 are statistically less than the WN
masses \citep[see][for a review]{c07}. For the WC and WO stars, there is growing evidence that they are both in the 
similar evolution phases of post helium burning, but at different surface temperatures and initial mass 
ranges \citep{sht12,me16}.
In our calculations, we only focus on the formation of WR+O binaries involving a WN star. 
%be a WN+O system after the mass transfer phase, the subsequent evolution involving WR wind mass loss
%is not followed.

\subsection{Examples of Binary Evolution}

To illustrate the detailed evolutionary sequences,
in Fig.~1 we show the evolutionary tracks of a binary in Model I with the initial parameters
of $ M_{\rm 1,i} =40M_{\odot}$ and $ q_{\rm i} = 2.0 $, and  the orbital periods
of 5 (top panel) and 40 days (bottom panel), corresponding to the Case A and Case B
mass transfer, respectively. The left, middle, and right panels show the evolution of the binary in the
Hertzsprung-Russell (H-R) diagram, the mass loss rate and the $ \Omega/\Omega_{\rm cr} $ value of the secondary, 
and the secondary mass and the orbital period, respectively.
Prior to the mass transfer, the primary has lost part of its mass due to the stellar wind,
which results in a slight widening of the binary orbit (in the blue curve).
In the top panel, the mass transfer starts at
the  age of $ \sim 3.9$ Myr, when the primary is still on the main-sequence. The mass transfer occurs at
a rate of several $10^{-3} M_{\odot}\,\rm yr^{-1}$ on the thermal timescale of the primary.
About $ 9M_{\odot} $ material is transferred in this rapid Case A phase, and both the components
roughly have similar mass of $ \sim27M_{\odot}$. After that the mass transfer rate declines to several
$ 10^{-6} M_{\odot}\,\rm yr^{-1}$. In this
slow Case A mass transfer phase, the mass transfer is driven by the nuclear evolution of the
primary. When core hydrogen is exhausted, the primary has a mass of $ \sim20M_{\odot} $.
It attempts to expand because the hydrogen shell ignites, leading to the Case AB mass transfer.
The mass transfer rate is $\sim10^{-4}-10^{-3} M_{\odot}\,\rm yr^{-1}$. Most of the 
hydrogen envelope is transferred during this phase, leaving a $14.2M_{\odot} $ helium burning star.
After mass transfer, the primary mass decrease to $12.6M_{\odot} $ only due to the WR stellar wind
(at a rate of $\sim 10^{-5} M_{\odot}\,\rm yr^{-1}$).
The orbital period finally grows to about 9 days. The secondary mass
(in the black curve) reaches to $ 31M_{\odot} $ by accreting $ 11M_{\odot} $ material during
the whole mass transfer phase. The $ \Omega/\Omega_{\rm cr} $ always $ \lesssim0.6 $ and 
has a large fluctuation, which can be low to $ \sim0.2 $ in the slow Case A mass transfer phase. 
The average mass transfer efficiency $ \beta_{\rm Av} $
is about 0.55. In the bottom panel, the primary fills its RL after the core hydrogen exhaustion at
an age of 4.7 Myr. The Case B mass transfer occurs rapidly at a rate of
$ 10^{-3}-10^{-2} M_{\odot}\,\rm yr^{-1}$, until the $ \sim17M_{\odot} $  hydrogen envelope
is stripped. The primary becomes a $ 16.2M_{\odot}$ helium burning star, and then a $ 15.3M_{\odot}$ 
WN star due to a stellar wind. The secondary increases its mass
to $ 22.4M_{\odot}$, and the final orbital period is 81.5 days. During the mass transfer, 
the $ \Omega/\Omega_{\rm cr} $ rapidly increases to be $ \sim0.9 $ and then roughly keeps this value to the end.
In this case the average mass transfer efficiency $ \beta_{\rm Av} \sim0.14$. 

Figure~2 shows the same evolutionary sequences for the binary but with Model II adopted.
Here half of the transferred mass is assumed to be accreted by the secondary. In the top panel,
the binary leaves a $ \sim14M_{\odot} $ WN star and a $ \sim29M_{\odot} $ O star with
$ P_{\rm orb} \sim9.5$ days. In the bottom panel, the resulting secondary star is much more massive
than that in Model I. After the mass transfer, the system possesses a WN+O binary with $ M_{\rm WN}=15.3M_{\odot} $, $ M_{\rm O}=27.9M_{\odot} $ and $ P_{\rm orb} \sim 77 $ days.

\subsection{Parameter distributions of WN+O binaries}

Transfer of mass and angular momentum in massive binaries can greatly influence
the properties of the formed WN+O system. Figures~3 and 4 show the calculated
mass $M_{\rm O}$ of the O star and the orbital period as a function of the mass $M_{\rm WN}$
of the WN star in Model I. The panels from top to bottom correspond
to the results with increasing mass ratio $ q_{\rm i} $. In each panel, the black curves
from left to right represent the cases with  the primary mass increasing from $ 25M_{\odot}$ to
$ 60M_{\odot}$ with an interval of $ 5M_{\odot}$. The squares in each of these curves indicate the WN+O binaries that have formed with different $ P_{\rm orb,i}$. The blue dashed curves are used to distinguish
Case A and Case B mass transfer. The two green dashed lines indicate the
mass ratios  $ M_{\rm WN}/M_{\rm O} = 0.5$ and 1.0. The red circles represent
the nine observed WN+O binaries with known binary parameters from the WR catalogue \citep{v01,rc15}.
Their basic parameters are listed in Table 1.

Fig.~5 presents the evolutionary tracks of the primary in the H-R diagram.
The left and right panels correspond to the binary systems with $ M_{\rm 1,i} =40M_{\odot} $ and increasing
initial orbital periods, and with $ P_{\rm orb,i} = 5 $ days and increasing initial primary masses, respectively.
The black and gray lines donate the primaries with the mass fraction of surface hydrogen 
$ X_{\rm surface} \geq 0.4$ and $ X_{\rm surface} <0.4 $, respectively.
The circle symbols denote the observed Galactic WN stars with detectable hydrogen, and the triangle symbols
refer to the hydrogen-free stars. The data of the WN stars with $ M_{\rm WN} <30M_{\odot} $ is taken \citep{h06,sht12}.
We can see that the modelled tracks can generally cross the observed distribution, the calculated WN stars seem to
have higher surface temperatures than observed, envelope inflation may be the reason for this discrepancy \citep{h06,y12,sht12}.
Here we assume that the calculated and observed WN+O binaries are in the similar evolutionary states, the parameter 
distribution of WN+O binaries can be used to constrain the progenitor evolution \citep[see][for an example]{e09}.

The formation of WN+O binaries depends on the stability of mass transfer and hence the initial mass ratio.
Generally the larger the initial mass ratios $ q_{\rm i}$,
the smaller the parameter spaces for stable mass transfer.
Systems with $ q_{\rm i} \gtrsim 2 $ may evolve to WN+O binaries
only when the initial binary orbital periods are larger than about 4 days \citep{sl14}.
Figures 3 and 4 show that the  WN/O mass ratios increase (up to $\gtrsim 1$) but the orbital periods decrease
with increasing initial mass ratio.
The minimum of the orbital periods is $ \sim2 $ days.

In Fig.~6 we present the calculated parameter distributions of WN+O binaries in Model II,
which requires the maximum initial mass ratio to be $ \sim2 $,  smaller than  in Model I.
The WN/O mass ratios can reach $ \gtrsim0.5 $ for systems with $ M_{\rm 1,i} \gtrsim 35M_{\odot}$.
The calculated orbital periods  generally $ \gtrsim 6 $ days.

Depending on the initial orbital periods, the binary systems may experience Case A or Case B
mass transfer in the evolution. In the former case, when the initial orbital period is longer, the primary can
develop a heavier core before the mass transfer, resulting in a more massive WN star.
In the latter case the produced WN mass is not sensitive to the initial
period, since the primary evolves as a single star during the core hydrogen burning phase without any
interaction with the secondary \citep[see also][]{p05}. This is clearly demonstrated in
Fig.~7, which shows the calculated WN mass as a function of the initial primary masses in Models I
(left) and II (right). While the masses of WN stars evolved from case A mass transfer significantly 
depend on the initial orbital periods, systems with Case B mass transfer tend to produce WN stars 
with similar masses. For latter binaries, we derive a relation between the masses of the initial primary stars
and the WN stars as follows
\begin{equation}
M_{\rm WN} = 0.46M_{\rm 1,i}-3.28,
\end{equation}
in Model I, and
\begin{equation}
M_{\rm WN} = 0.47M_{\rm 1,i}-3.87.
\end{equation}
in Model II. These results are close to the relation given by \citet{wl99} for the systems
with quasi-conservative Case B mass transfer.

Fig.~8 shows the average mass transfer efficiencies $ \beta_{\rm Av} $
as a function of the orbital periods of the WN+O binaries in Model I. We use different symbols
to indicate the binaries with different initial parameters. The blue dashed curve distinguishes Case A and
Case B mass transfer. We can find that $ \beta_{\rm Av} $ decreases with increasing
orbital period, in the range of $\sim 0.2-0.7 $ for Case A mass transfer
and $\lesssim 0.2$ for Case B mass transfer. Mass accretion tends to spin up
the secondary, while the tidal interactions in close binaries attempt
to synchronize the spin with the binary orbit. So systems with shorter orbits have higher
mass transfer efficiency, and the mass transfer in the binaries with wider orbits are less efficient.

\section{Discussion}

%For the selected WR+O binaries with known mass ratios and orbital periods, the systems WR 31
%and WR 97 have obviously low masses which inferred from the orbital inclinations \citep{v01},
%however the masses of O stars are much lower than the expected spectroscopic masses of
%$ \sim20-30 M_{\odot}$ \citep{l96}. Since the WR  mass of WR97 is just $ \sim2.3 M_{\odot}$,
%we take the minimal masses of primary stars as $ 10M_{\odot} $.

Our calculations demonstrate the possible distribution of the parameters of WN+O binaries.
In Model I the binaries contain a WN star of mass  $\sim6-25 M_{\odot}$ and an OB
main sequence star of mass $\sim7-54M_{\odot}$ with orbital period ranging from
about 3 days to several hundred days (see Fig.~3 and 4). In Model II,
the binaries contain  a WN star of mass $\sim6-25M_{\odot} $ and an O star of mass
$ \sim20-50M_{\odot} $ with orbital period longer than $\sim6$ days (see Fig.~6).

These results can be compared with observed WN+O binaries to constrain the input parameters
and mechanisms adopted. We first note that
conservative evolution produces WN+O systems with extremely
low mass ratios ($ M_{\rm WN}/M_{\rm O}\lesssim0.3$) and orbital
periods  much longer than observed \citep{w01}. Non-conservative mass transfer 
can effectively decrease both
the masses of the O stars and the orbital periods via mass and angular momentum loss.
The average mass transfer efficiency in Model I varies with a large range. It
can reach $\sim0.7 $ for Case A mass transfer, because tidal interactions spin
down the secondary so it can keep accreting a substantial amount of the mass. For case B
mass transfer, $ \beta_{\rm Av} $ is lower than about 0.2 because the tidal effect
is negligible in wide binaries. So the secondary can keep rapid rotation after being spun up during
the initial mass accretion phase, and most of the transferred material is ejected out of the binary.
In the $ M_{\rm WN}-M_{\rm O} $ plane of Figs.~3 and 4, all of the observed WN+O binaries can
be well covered with $ q_{\rm i}\sim 2.0-2.5 $ in model~I.
In Model II the modeled O stars are considerably  more massive than that in Model I and observations.

The mass transfer efficiency can be strongly constrained by the parameters of the binary WR 35a.
The estimated masses of the binary component are $\sim18M_{\odot} $ for the WN star and
$\sim19M_{\odot} $ for the O star, and the orbital period is 41.9 days \citep{g14}.
From Eq.~(2) or (3), we can infer that the initial mass of the primary star should be
$ >46M_{\odot} $, so $ q_{\rm i}\gtrsim2.5 $. The WN/O mass ratio of  $ \sim0.8-1.3 $ \citep{g14}
disfavors Model II for the formation of such WN+O  binaries, implying
that the accreted mass by the O star is very small. During Case B mass transfer in Model I most
of the transferred material can be expelled, leading to binaries with relatively large WN/O mass ratios.
So the progenitor binary is very likely to have experienced highly non-conservative mass transfer.

Our results are compatible with previous studies.
\citet{p05} investigated the formation of WR 21, WR 127 and WR 153ab. They
pointed out that these systems can be produced only if the mass transfer is highly non-conservative,
with  only $ \sim10\% $ of the transferred mass being retained by the mass gainer.
In our calculations, these binaries can be formed in Model I with inefficient mass transfer
if the initial mass ratios $ q_{\rm i} \lesssim2.5$ and the mass transfer efficiency is $\gtrsim0.2$.
Binary evolution in Model II may  produce such binaries only when the $ q_{\rm i} \sim2$.
\citet{d07} suggested that binaries with smaller initial orbital periods evolve more conservatively
than wider systems, which is also in agreement with our results of Model I (see Fig.~8).

The systems WR 151 and WR 155 have the large WN/O mass ratios of $\sim 0.7-0.8 $ and the short orbital periods
of $\sim1-2 $ days, these two binaries may be formed through unstable Case C mass transfer. 
Alternatively we only assume that the initial binary systems have circular orbits, the mass transfer 
in highly eccentric binaries may also contribute to produce these short periods \citep{dk16a,dk16b}.
In some binaries, the O stars
have been observed to rotate faster than synchronous rotation \citep{m81,u88,b90,m94,s15},
suggesting that the progenitors have
experienced stable mass transfer rather than contact phases or common envelope evolution.
Further observations on the spins of the O stars in massive binaries \citep[e.g.,][]{d13}
can help understand the formation of WN+O binaries.

In summary, our calculations demonstrate that massive binary evolution with
rotation-dependent mass transfer model can satisfactorily match the observations of WN+O binaries.
%provided that there is efficient angular momentum loss along with mass loss.

\acknowledgements
We thank the referee for her/his helpful suggestions that improved this paper.
This work was supported by the Natural Science Foundation of China
(Nos. 11603010, 11133001, 11333004 and 11563003), the Strategic
Priority Research Program of CAS (Grant No. XDB09000000), the Natural Science
Foundation for the Youth of Jiangsu Province (No. BK20160611), and the Fundamental 
Research Funds for the Central Universities (No. 020114380015).

%\end{acknowledgements}

\clearpage

\begin{figure}

\centerline{\includegraphics[scale=0.5]{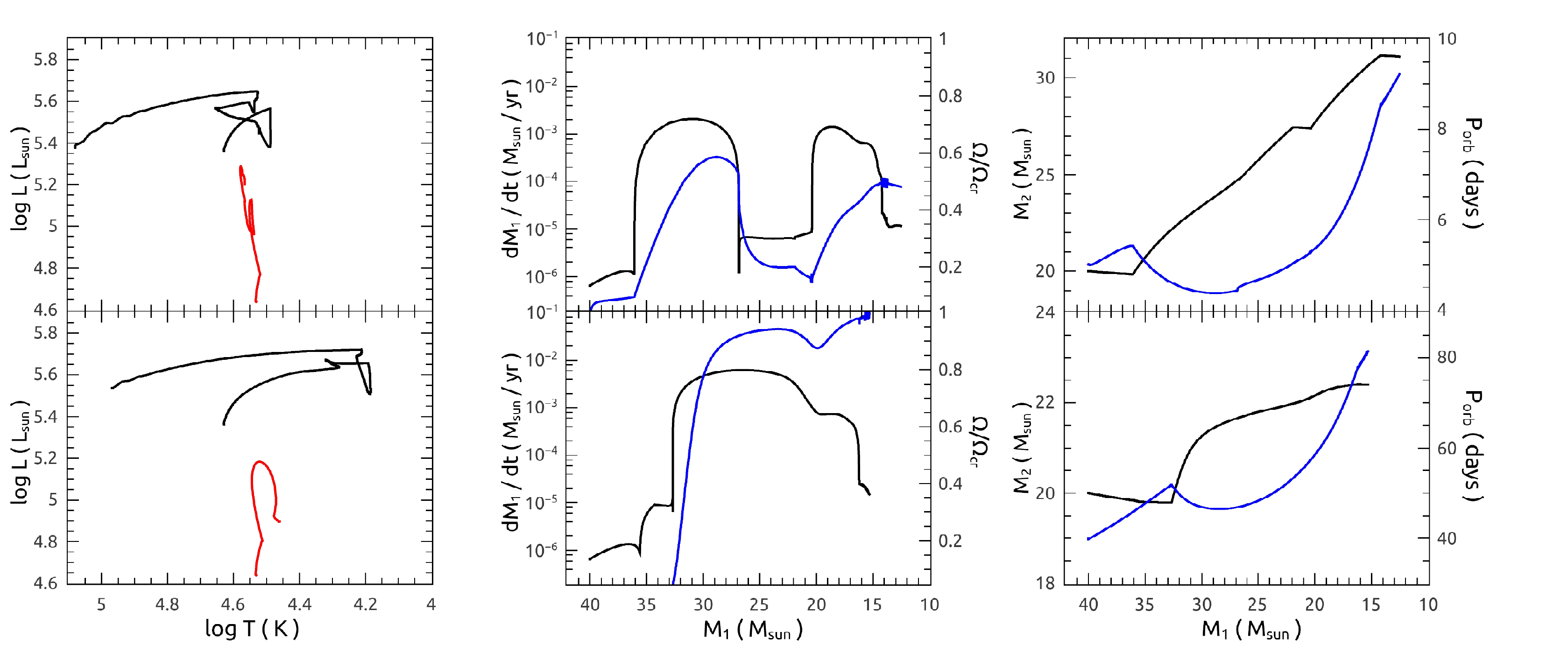}}
\caption{Evolution of massive binaries with $ M_{\rm 1,i}=40M_{\odot} $, $ q_{\rm i}=2.0 $,
and $ P_{\rm orb,i} =$ 5 days (top) and 40 days (bottom) in Model I. The evolutionary sequences
of the primary (black) and the secondary (red) in the H-R diagram are presented in the
left panels, the mass loss rates (black) and the $ \Omega/\Omega_{\rm cr} $ values (blue)
in the middle panels,  the secondary masses (black)
and the orbital periods (blue) in the right panels.
   \label{figure1}}

\end{figure}

\begin{figure}

\centerline{\includegraphics[scale=0.5]{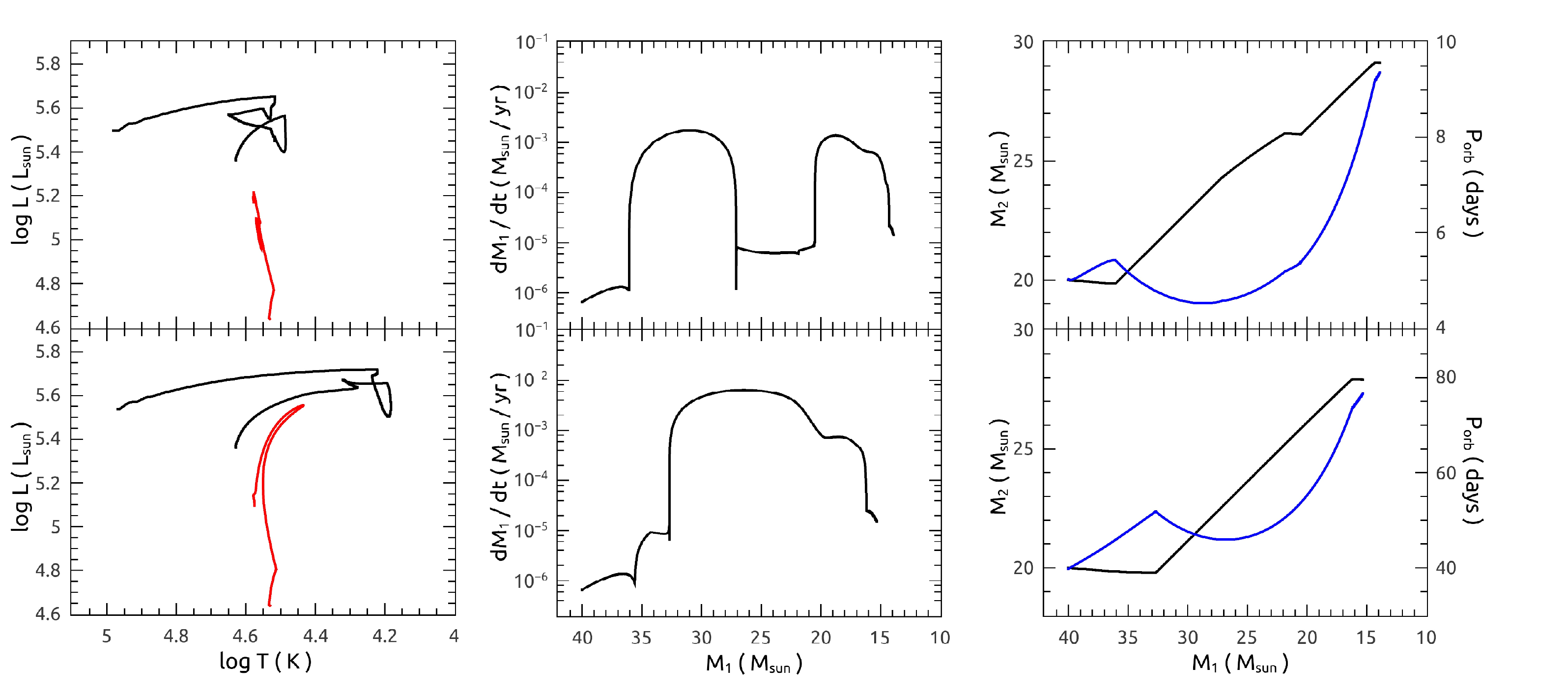}}
\caption{Similar to Fig.~1 but in Model II.
 \label{figure2}}

\end{figure}

\begin{figure}

\centerline{\includegraphics[scale=0.6]{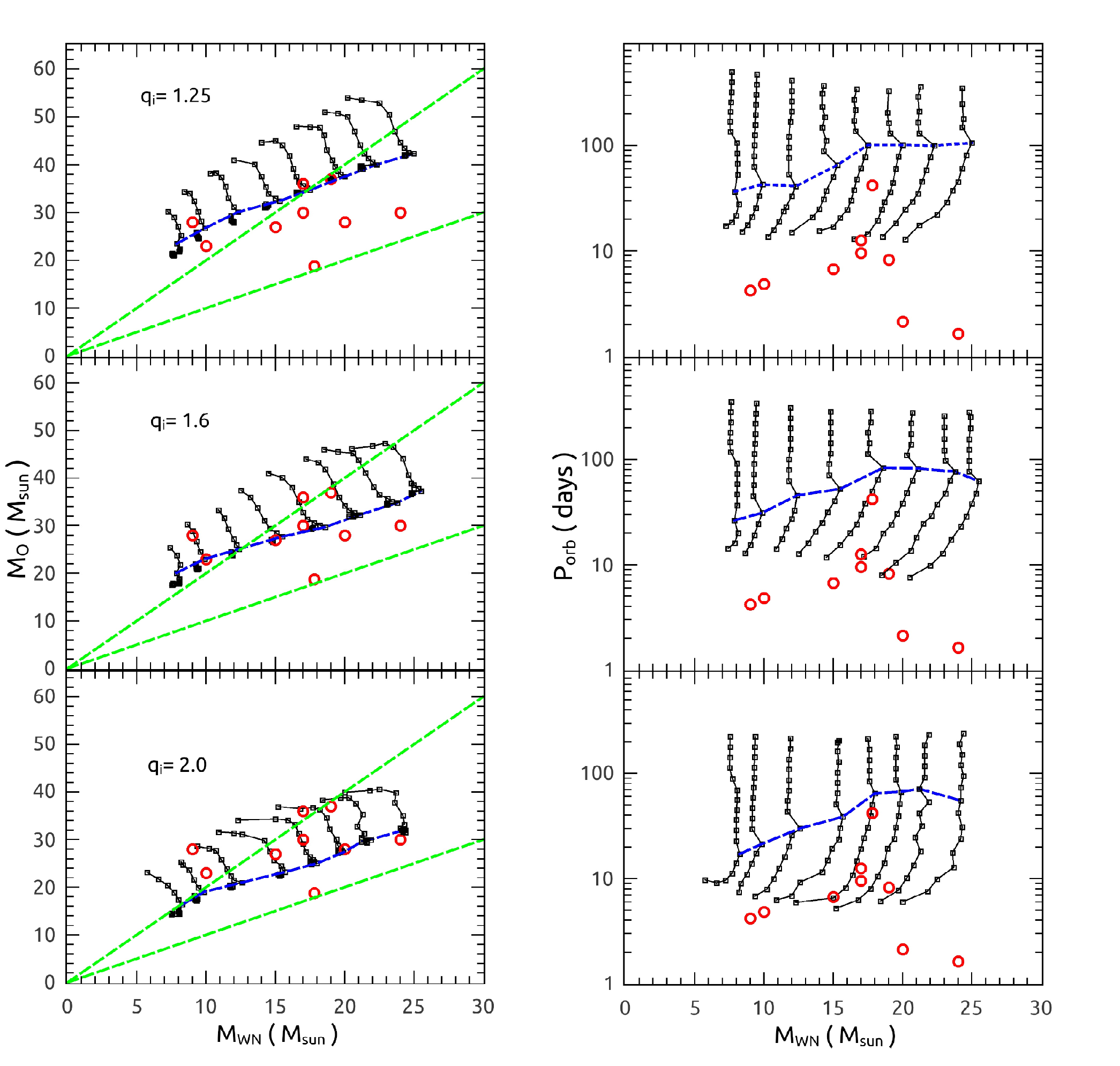}}
\caption{The calculated distributions of WN+O binaries are shown in
the $ M_{\rm WN}-M_{\rm O} $ (left) and $ M_{\rm WN}-P_{\rm orb} $ (right) planes.
The initial mass ratios of progenitor systems are taken to be
1.25 (top), 1.6 (middle) and 2.0 (bottom). Every square gives the position
of derived WN+O binary, after the mass transfer, in the parameter spaces.
Squares in each black curve correspond to the progenitors with the same
initial primary mass but different initial orbital period, the closest
curves from left to right mean that the initial primary mass has an increasing interval
of $ 5M_{\odot} $. The blue dashed curves are used to distinguish the
Case A (left panel: above; right panel: below) and Case B (left panel: below; right panel: above) 
binaries during the mass transfer phases. The two green dashed
lines correspond to mass ratios $ M_{\rm WN}/M_{\rm O} = $ 0.5 and 1. The
red circles show the positions of the nine observed WN+O binaries
with $ M_{\rm WN} <30M_{\odot} $ \citep{v01,rc15}.
\label{figure3}}

\end{figure}

\begin{figure}

\centerline{\includegraphics[scale=0.6]{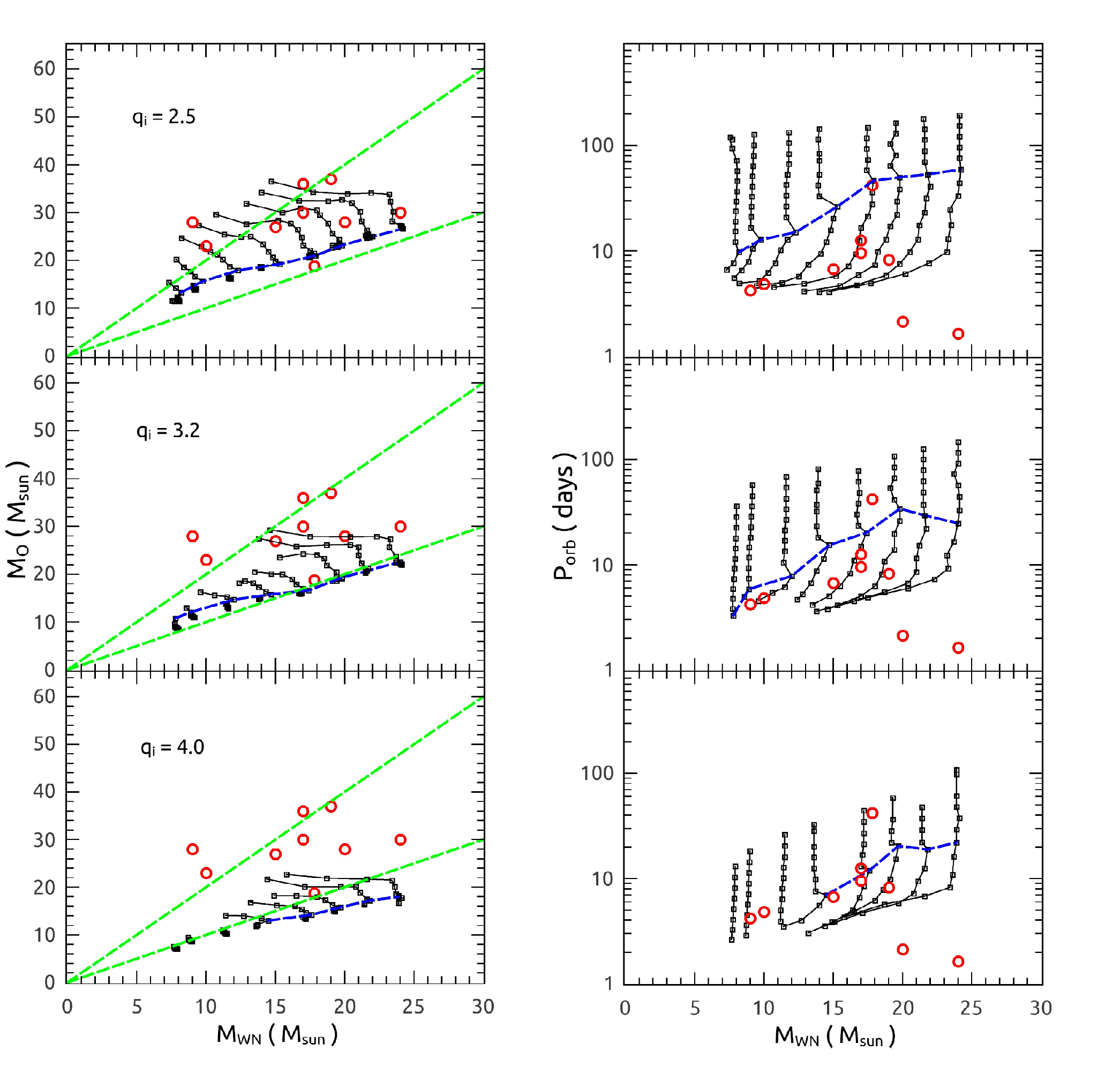}}
\caption{Similar to Fig.~3, but with $  q_{\rm i} $ = 2.5 (top), 3.2 (middle) and
4.0 (bottom), respectively.
\label{figure4}}

\end{figure}

\begin{figure}

\centerline{\includegraphics[scale=0.6]{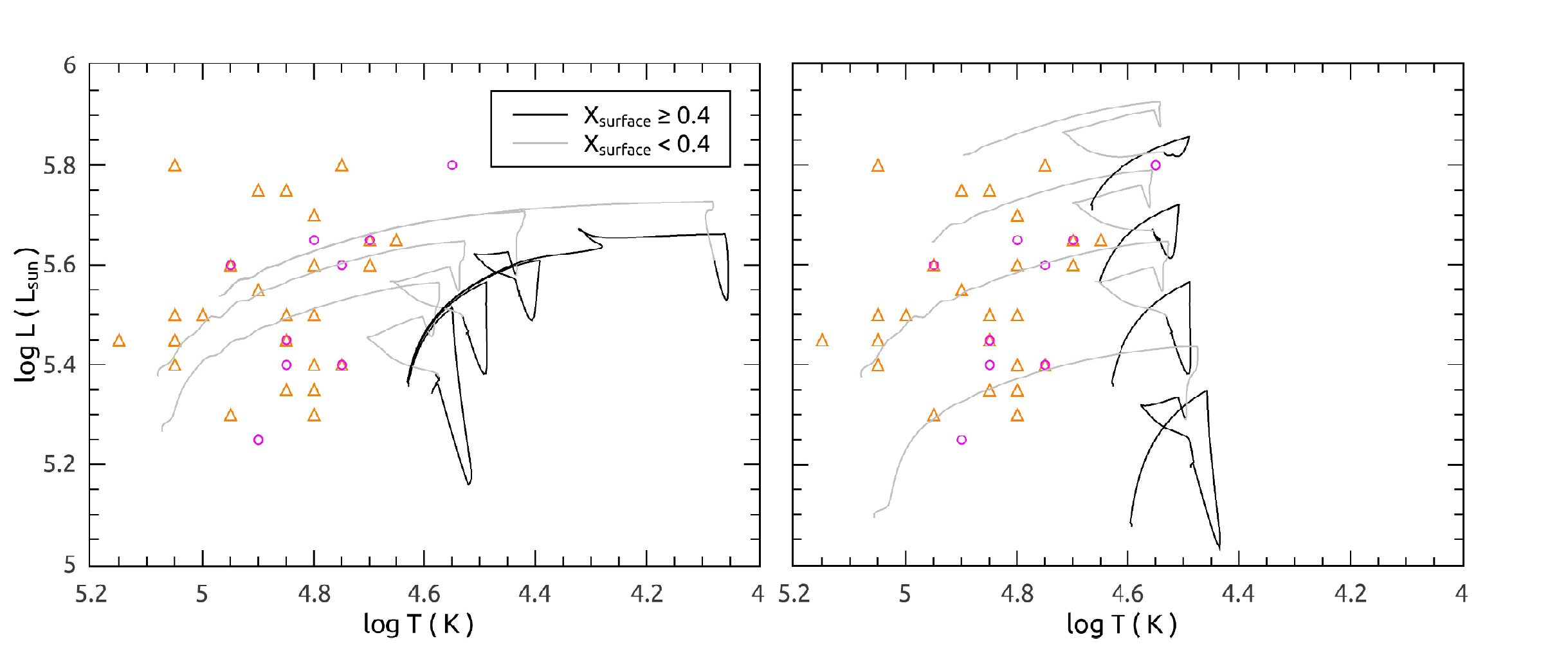}}
\caption{Hertzprung-Russell diagram for the primaries in massive binary systems.
The left panel corresponds to the binaries with $ M_{\rm 1,i} =40M_{\odot} $ and $ P_{\rm orb,i} = 3 $,
5, 10, 100 days (from bottom to top) and the right panel donates
the binaries with $ P_{\rm orb,i} = 5 $ days and $ M_{\rm 1,i} =30$, 40, 50, 60$M_{\odot}$  
(from bottom to top). The circle and triangle symbols refer to the observed Galactic WN stars,
corresponding to the stars with detectable hydrogen and the hydrogen-free stars, respectively. 
\label{figure5}}

\end{figure}

\begin{figure}

\centerline{\includegraphics[scale=0.6]{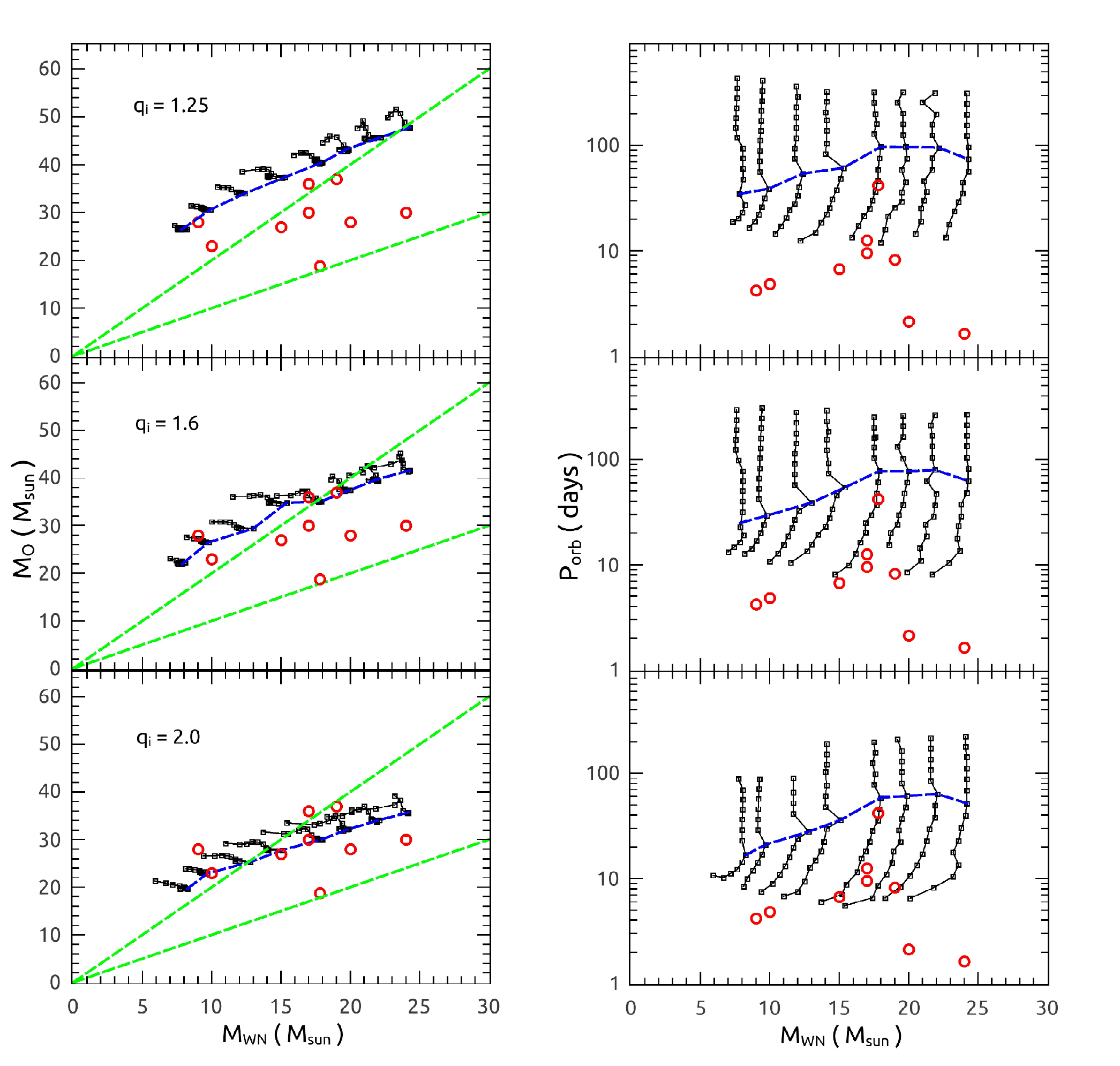}}
\caption{Same as Fig.~3, but Model II is adopted. The corresponding maximal mass
ratio is $ \sim2 $.
\label{figure6}}

\end{figure}

\begin{figure}

\centerline{\includegraphics[scale=0.6]{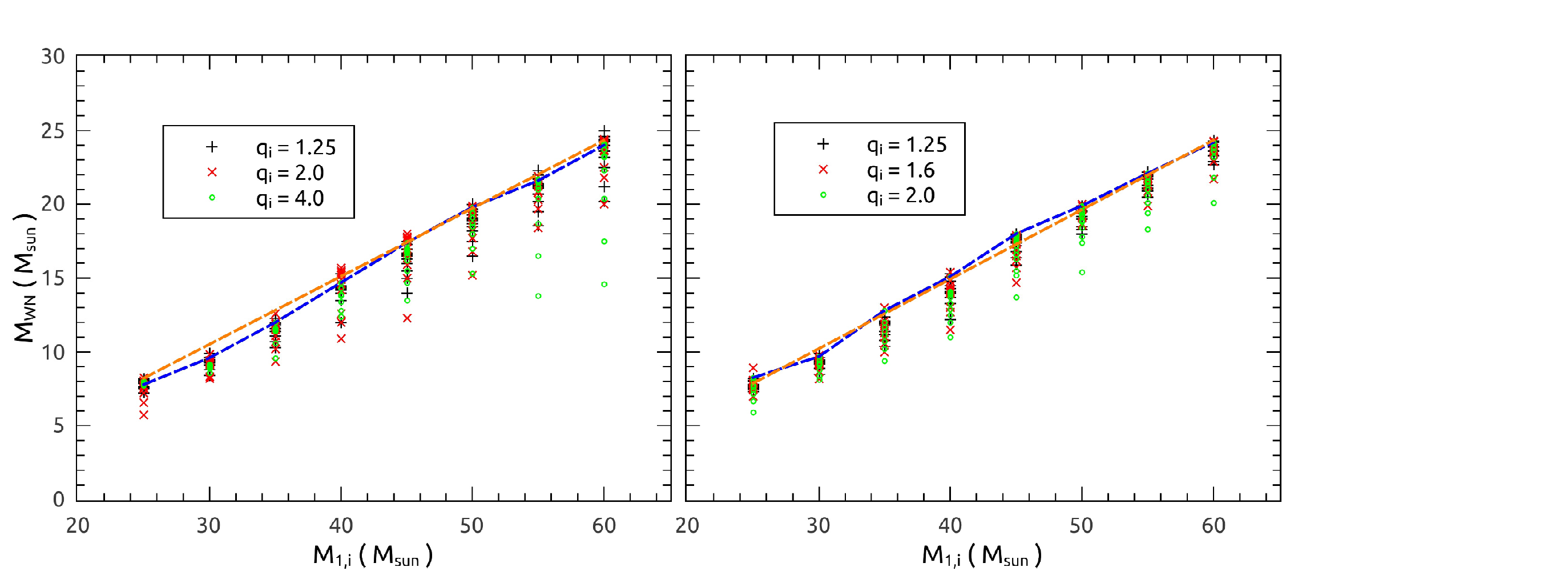}}
\caption{Relation between the initial primary mass ($ M_{\rm 1,i} $) and the WN mass ($ M_{\rm WN} $) in Model I
(left) and II (right). Different signals correspond to the binaries with different mass ratios. The same 
signals for a special primary mass, from bottom to top, correspond to the increasing initial orbital periods.
The blue dashed line is used to distinguish the Case A (below) and Case B (above) binaries, and the orange dashed line  
gives the relation as a linear fit for Case B binaries.
\label{figure7}}

\end{figure}

\begin{figure}

\centerline{\includegraphics[scale=0.6]{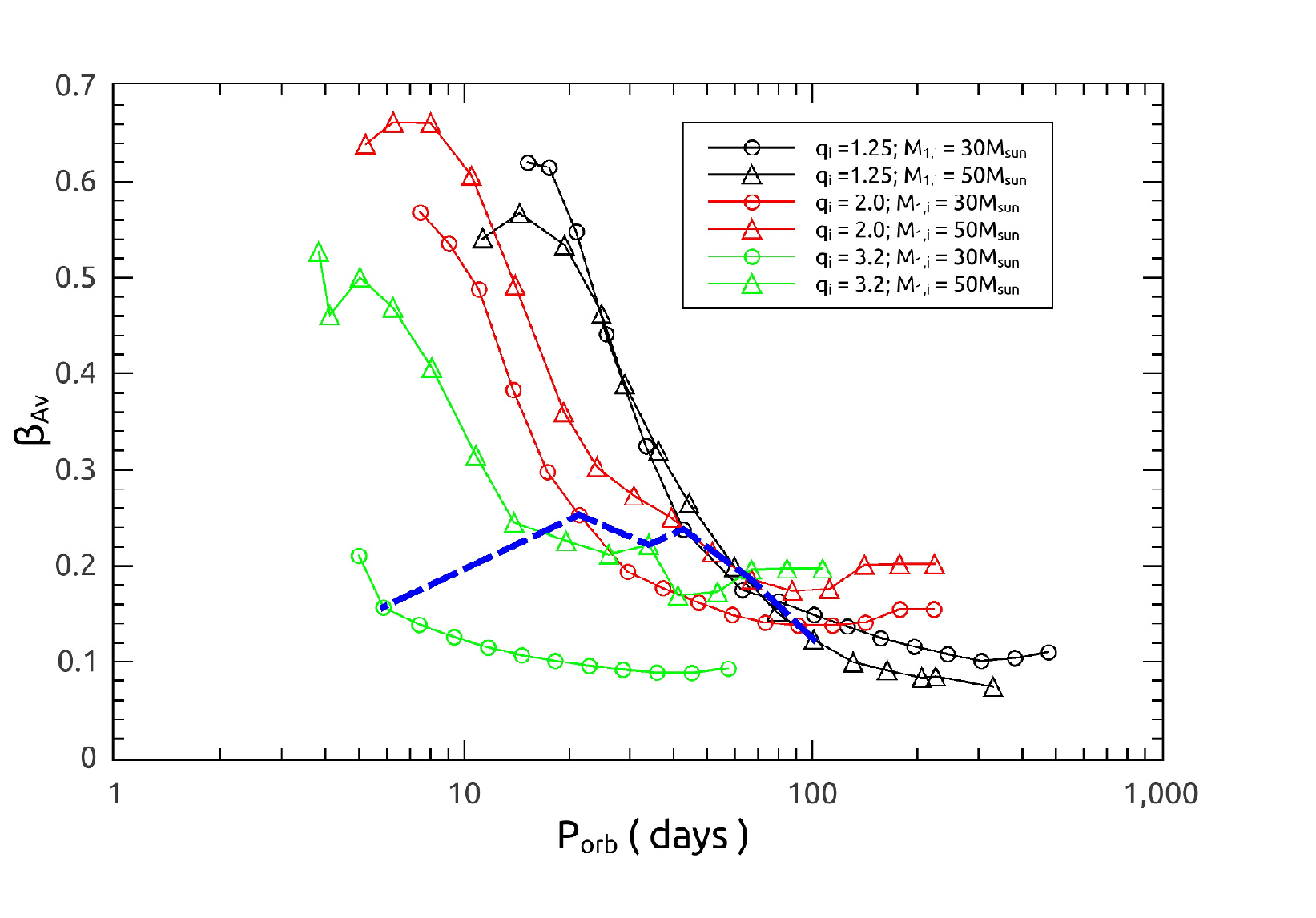}}
\caption{Average mass transfer efficiencies $ \beta_{\rm Av} $, in Model I,
as a function of the orbital periods of calculated WN+O binaries.
Every signal corresponds to one
individual calculated result. Different binary parameters are marked
with different colors and signals.
The blue dashed curve is used to distinguish the results from Case A (above) and Case B
(below) systems.
\label{figure8}}

\end{figure}

\clearpage

\begin{table}
\begin{center}
\caption{Basic parameters of selected WN+O binaries from the WR catalogue
\citep{v01,rc15}.\label{tbl-1}}
\begin{tabular}{lcccccc}
\hline
WR number &  HD/name & Spectral Types & $P_{\rm orb} $ (days)& $M_{\rm WN}(M_{\odot})$ &$M_{\rm O}(M_{\odot})$&
$  M_{\rm WN}/M_{\rm O}$   \\

\hline
WR 21  &HD 90657  & WN5+O4-6  & 8.25& 19 & 37 & 0.52 \\
WR 31$^{*}$  &HD 94546  & WN4+O8V   & 4.83& 10 & 23 & 0.43 \\
WR 35a &SMSP 5    & WN6+O8.5V & 41.9& 18 & 19 & 0.95 \\
WR 97$^{*}$  &HD 320102 & WN5+O7    & 12.6& 17 & 30 & 0.56 \\
WR 127 &HD 186943 & WN3+O9.5V & 9.56& 17 & 36 & 0.47 \\
WR 139 &HD 193576 & WN5+O6III-V& 4.21& 9 & 28 & 0.34 \\
WR 151 &CX Cep    & WN4+O5V  & 2.13& 20 & 28 & 0.71 \\
WR 153ab &HD 211853 &WN6+O6I & 6.69& 15 & 27 & 0.54 \\
WR 155 &CQ Cep    & WN6+O9II-Ib& 1.64& 24 & 30 & 0.8 \\

\hline
\end{tabular}
\end{center}
*-The estimated masses of binary components, from the catalogue, are extremely low for their
spectral types (4 and $ 9M_{\odot} $ for WR 31;
2.3 and $ 4.1M_{\odot} $ for WR 97). We use the spectral 
masses of the O stars suggested by \citet{l96} and derive the WN masses with the mass ratios.  
\end{table}

\end{document}